# Heavy quark effective theory for charm mesons

S. Singh*, M. Batra , A. Upadhyay
*School of Physics and Material Science, Thapar University Patiala 147004, Punjab, INDIA*
* email: satti376@gmail.com

## Introduction

We are interested in the Mesons with one heavy quarks and one light anti-quark out of up, down or strange quark. Such systems are of much interest because in such systems we encounter new symmetries which are not present even in the original theory (QCD). Such symmetries have been exploited to formulate a theory that describes the low energy interactions among the heavy mesons. We studied the masses of even and odd parity charmed mesons within the framework of heavy quark effective theory and the chiral perturbation theory. Various approaches have been used in literature to study heavy-light mesonic sector in literature.  Specifically effective field theories exploit the symmetries of hadrons(mesons) to make model independent predictions when the dynamics of these hadrons are too hard to solve explicitly. For example, the properties of a hadron containing a very heavy quark are insensitive to the orientation of the heavy quark spin. This heavy quark spin symmetry can be used to make predictions for the production and decay of heavy mesons and quarkonia at collider experiments.

## Spin and Flavor Symmetries

In heavy-light systems the light quarks are bound to the heavy quark with soft gluons which do not have sufficient energy to probe the heavy quark. Compton wavelength, of heavy quark $(\lambda_Q \propto 1/m_Q)$ is very-very small as $m_Q \rightarrow \infty$. Light quark can probe only distances much greater than $m_Q$ an hence cannot feel spin and flavor of the heavy quark [1].Here is the origin of new symmetries called spin and flavor symmetries. We can use these symmetries to relate the properties of various mesons differing in spin or flavor or both. If in a meson, with bottom quark as heavy quark,bottom is replaced by a charm quark with same or different spin, then resulting meson will have similar properties. When bottom quark is changing into charm quark a spontaneous symmetry is broken and Goldstone Boson (Π,η or K-meson) are exchanged between the fields representing the mesons.

## Effective Lagrangian and Mass Formula

Interaction between various mesonic states  is seen through the interaction of fields. To study these dynamics an effective Lagrangian [2] is formulated making the use of above symmetries in which effects of heavy quark fields are integrated out partly as they do not affect the dynamics to much extent due to above symmetries. Mass formula is then used for ground state $J^P = 0^- \& 1^-$ and first excited state $J^P = 0^+ \& 1^+$ charmed mesons up to one loop chiral corrections[3]. The masses are depending on a number of factors, some of which are symmetry breaking ( $\Delta_H, \Delta_s, \Delta_H^\sigma, \Delta_H^{(a)}$, $\Delta_S^\sigma, \Delta_S^{(a)}$ ) and others are symmetry conserving parameters ($\sigma_s, \sigma_H, a_s$ and $a_H$). The values of some parameters are poorly known or unknown & hence their contributions are absorbed in the other parameters. With resulting eight equations, we tried to fit the masses by varying the parameters to sufficient ranges to get the better central values for the masses of the charmed mesons which are ambiguous at various experiments. Similar work has been done in some research papers [4]. The possible range of some parameters is available from various experiments. For example possible ranges of different coupling constants $g, g'$ and $h$, are determined from various decay processes. An analysis of $D^*$ decay using one loop calculation without explicit excited states yields $g = 0.27^{+0.06}_{-0.03}$ [30]. From the widths of the non-strange resonances observed by Belle we have extracted  $h = 0.69 \pm 0.009$[31].  Both the coupling constants are of the order of unity. Taking an indication from here, in our work we vary these vary their values over the range 0-1 so as to include all possible values.

## Results

We worked at the energy range ~ 1 GeV. The masses of up and down quarks are taken to be equal and ~ 4 MeV, while the mass of strange quark is taken to be ~ 90 MeV.

The values of various parameters giving better values of masses are

$g = 0.01, g' = 0.01, h = 0.05, \delta_H = 4$,
$\delta_S = 432, \Delta_H = 14, \Delta_S = 128, a_H = 1.1$,
$a_S = 0.21, \Delta_H^a = -0.04, \Delta_S^a = 0.14$.

The comparison of calculated masses in our work and the values of masses at experiments is shown in the table 1.

We also studied the variation of mass splittings and spin splittings w.r.t various symmetry conserving and symmetry breaking parameters graphically, varying a particular

| Mesonic state | Experimental masses | Calculated masses |
|---|---|---|
| $D^{0,+}$ | 1867.21 | 1867.04 |
| $D_s^\pm$ | 1968.47±0.33 | 1967.74 |
| $D_0^0$ | 2318±29 | 2314.09 |
| $D_{s0}^+$ | 2317.8±0.6 | 2314.55 |
| $D^{*0,+}$ | 2008.60 | 2017.24 |
| $D_s^{*+}$ | 2112.1 | 2109.50 |
| $D_1^0$ | 2438±31 | 2433.52 |
| $D_{s1}^+$ | 2459.5±0.6 | 2456.29 |

**Table 1:** Experimental Vs. calculated masses

parameter keeping all other parameters fixed and watching the variation in splitting values. One such plot showing large variation in the mass splittings with the variation in symmetry conserving parameter $a_H$ is shown in Fig. 1. The terms proportional to $a_H$ results in $SU(3)_V$ violating mass splittings amongst the vector mesons. From these type of graphs we get an idea about the possible ranges of the parameters to be considered while calculations. After analyzing the graphs, a constrain fit has been applied to get appropriate values of the splittings. The parameter set satisfying these splitting values is

$g = 0.01, g' = 0.01, h = 0.01, \delta_H = 4$,
$\delta_S = 432, \Delta_H = 144, \Delta_S = 126, a_H = 1.1$,
$a_S = 0.2, \Delta_H^a = -0.03, \Delta_S^a = 0.14$.

The calculated mass splittings are :

$D_s^* - D_u^*(1^- - 1^-) = 94.0396$ MeV

$D_s - D_u(0^- - 0^-) = 96.578$ MeV

$D_s^* - D_u^*(1^+ - 1^+) = 20.3321$ MeV

$D_s - D_u(0^+ - 0^+) = 8.1228$ MeV

Similarly spin splittings have also been calculated and the results are in good agreement with the experimental values.

## References

[1] Matthias Neubert, hep-ph/10512222 v1
[2] E. Jenkins, hep-ph/9212295v1 21 Dec 1992
[3] T. Mehen and R. P. Springer, Phys.Rev. **D72** (2005) 034006.
[4] B. Ananthanarayan, S. Banerjee, K.Shivaraj, A. Upadhyay, Phys.Lett.**B651**:124-128,2007

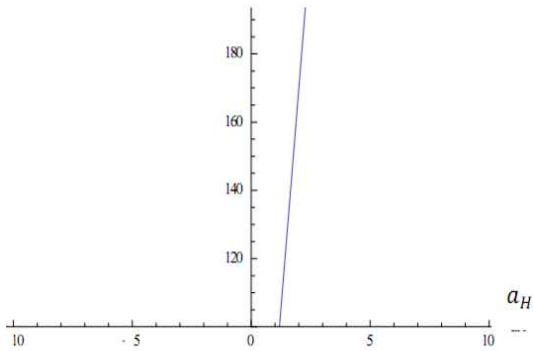